\documentclass[12pt]{article}

\usepackage{hyperref}
\usepackage{graphicx}
\usepackage{natbib}
\usepackage{multirow}
\usepackage{colortbl}

\title{Microwave emissivity of freshwater ice\\ Part II: Modelling the Great Bear and Great Slave Lakes}

\author{Peter Mills \\ Peteysoft Foundation\\\href{mailto:petey@peteysoft.org}{petey@peteysoft.org}}

\begin{document}

\maketitle

\begin{abstract}
Lake ice within
three Advanced Microwave Scanning Radiometer on EOS (AMSR-E) pixels
over the Great Bear and Great Slave Lakes have been simulated
with the Canadian Lake Ice Model (CLIMo).
The resulting thicknesses and temperatures were fed to a radiative transfer-based 
ice emissivity model and compared to the satellite measurements at
three frequencies---6.925 GHz, 10.65 GHz and 18.7 GHz.
Excluding the melt season,
the model was found to have strong predictive power, returning a correlation
of 0.926 and a residual of 0.78 Kelvin at 18 GHz, vertical polarization.
Discrepencies at melt season are thought to be caused by the presence of
dirt in the snow cover which makes the microwave signature more like
soil rather than ice.
Except at 18 GHz, all results showed significant bias compared to measured
values.
Further work needs to be done to determine the source of this bias.
\end{abstract}

\section{Background}

In \citet{Mills2012}, radiative transfer simulations of freshwater
ice---lake ice and Antarctic icepack---were compared with measurements from
the Advanced Scanning Radiometer on EOS (AMSR-E).
In the previous study, the validation of the ice emissivity model over lake
ice was rather crude: brightness temperatures (Tbs) were modelled for three
different ice thickness using a constant temperature.
These were compared with averaged AMSR-E measurements collected over the
entire winter season and separated from the open water measurements
using a clustering algorithm.
Since the stated purpose of the past study was to rigorously validate a
radiative transfer-based ice emissivity model using ice with easy-to-predict
dielectric constants (fresh water ice) and to do so on a point by point basis,
that is, radiance measurements are matched by location and time with 
the ice properties used to model the radiances, for lake ice, this has
not been fulfilled.

In this new study, we generate three radiance time series over the Great Bear
and Great Slave Lakes based on ice thicknesses
and temperature profiles from a thermodynamic lake ice model.
In \citet{Kang_etal2010}, these same results were used to statistically
relate ice thickness to AMSR-E brightness temperature, highlighting
the possibility for satellite ice thickness retrieval, 
which over saline water bodies, is still an unsolved problem.
It would be valuable to provide the relationship with a physical basis.
The three study locations, two on the Great Slave Lake---one near Yellowknife,
and one near the mouth of the Hay River, and one in the Great
Bear Lake, are AMSR-E pixels located wholly over
open water, although there is some doubt about the 6 GHz channels which
have a larger footprint \citep{Kang_etal2010}.

The thermodynamic ice growth model, the Canadian Ice Growth Model
(CLIMo), is described in \citet{Duguay_etal2003}.
\citet{Mills_Heygster2011} provide a much simpler, but nonetheless conceptually similar example of how to model ice growth.
\citet{Cox_Weeks1988} also model sea ice growth using such a thermodynamic
model.
On the basis of heat flux between water and atmosphere, determined by
the atmospheric state, ice growth can thereby be predicted.
In addition, CLIMo also simulates the circulation beneath the ice as well as the
presence of snow cover, which will act as an insulation barrier.

The ice emissivity model is briefly described in Part I \citep{Mills2012},
Section 2.1, and in full depth in \citet{Mills_Heygster2009}.
The snow is modelled in the same manner as the Antarctic icepack, 
using the mixture model for spherical inclusions from \citet{Sihvola_Kong1988}
assuming an ice volume-fraction of 40 \%.
The snow depth was not modelled, but rather measured by nearby weather
stations.
Snow-air, air-ice and ice-water interfaces were modelled as discontinous,
that is, reflecting, while the ice within the sheet was assumed to be
continous, that is, non-reflecting.

\section{Results and discussion}

\begin{figure}
\begin{center}
\includegraphics[width=.9\textwidth]{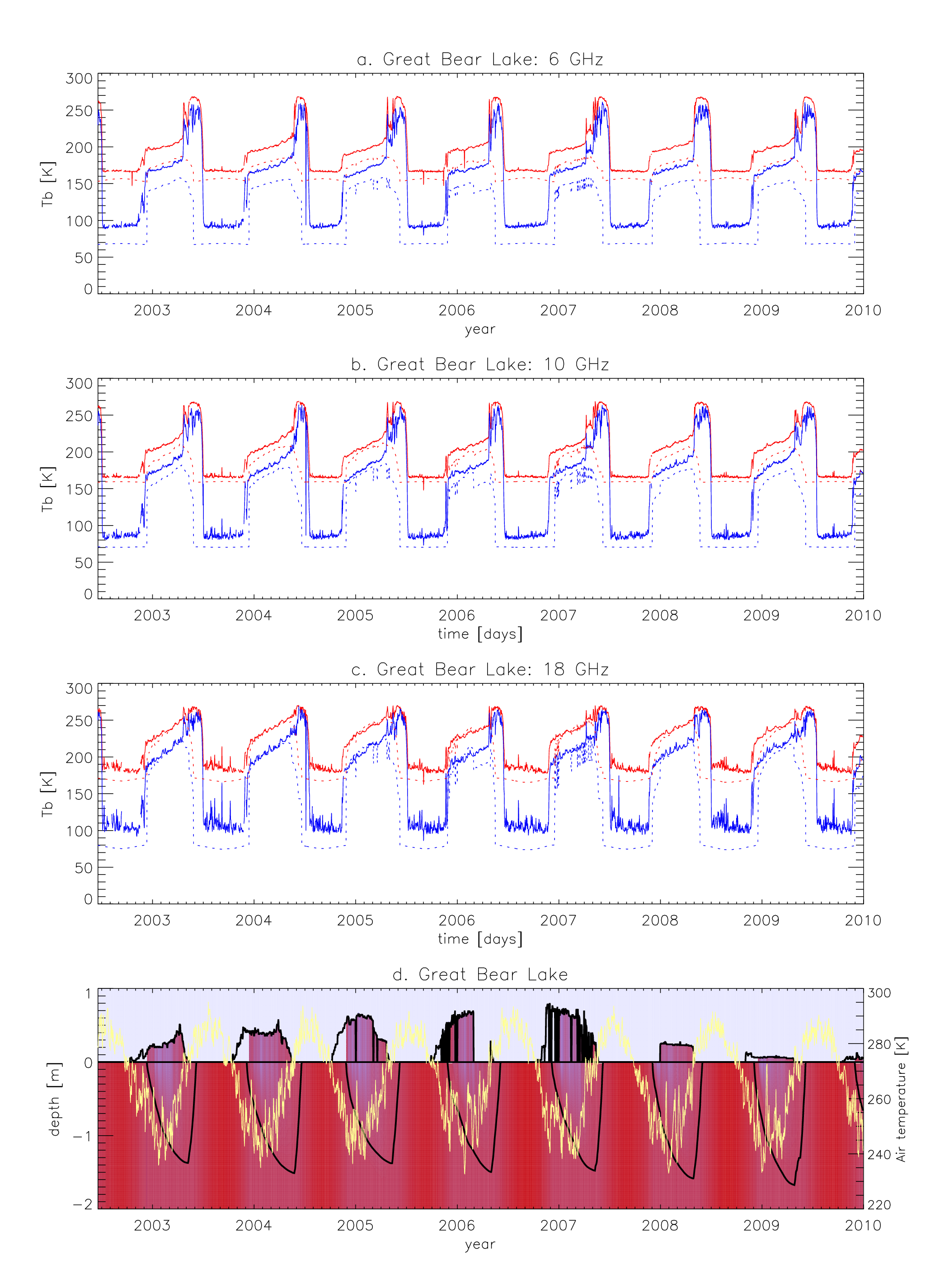}
\includegraphics[width=0.6\textwidth]{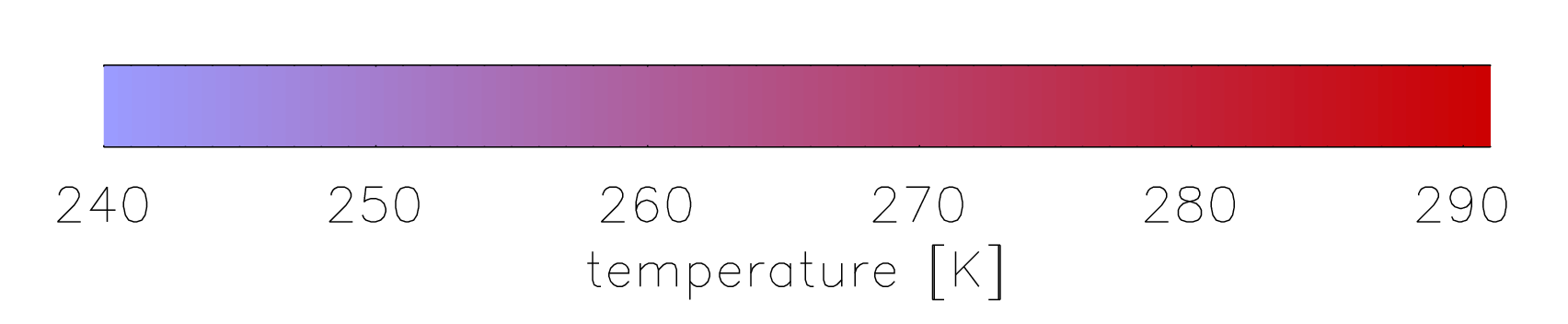}
\caption{(a), (b) and (c): Time series of measured versus modelled brightness temperatures
over the Great Bear Lake, 6, 10 and 18 GHz, respectively.
(d): Ice thickness (black lines), snow depth (black lines), ice and water temperature (shading) and air temperature (yellow line).}
\label{GBL_tbs}
\end{center}
\end{figure}

\begin{figure}
\begin{center}
\includegraphics[width=.9\textwidth]{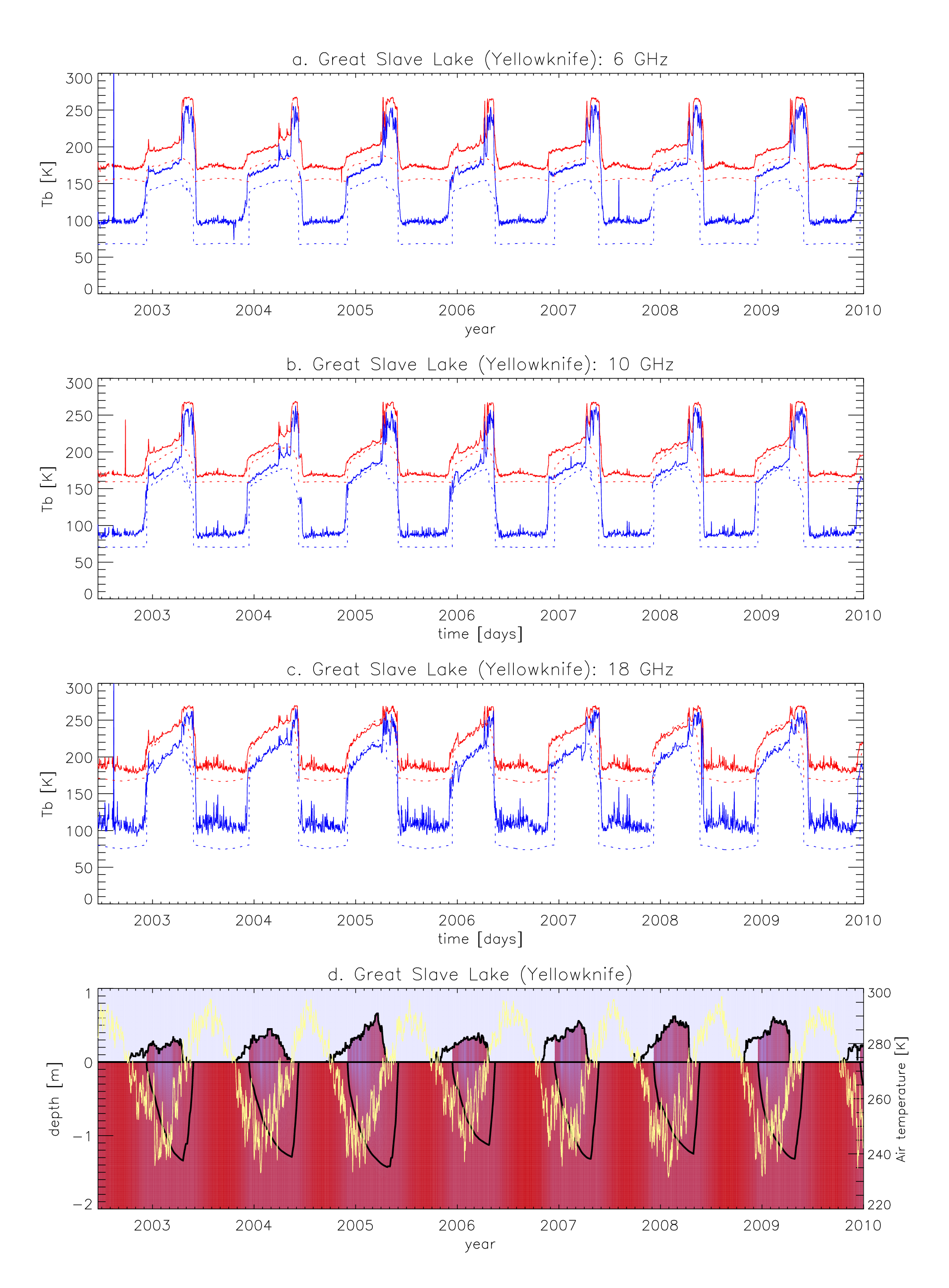}
\includegraphics[width=0.6\textwidth]{GBSL_legend}
\caption{(a), (b) and (c): Time series of measured versus modelled brightness temperatures
over the Great Slave Lake near Yellowknife, 6, 10 and 18 GHz, respectively.
(d): Ice thickness (black lines), snow depth (black lines), ice and water temperature (shading) and air temperature (yellow line).}
\label{GSL1_tbs}
\end{center}
\end{figure}

\begin{figure}
\begin{center}
\includegraphics[width=.9\textwidth]{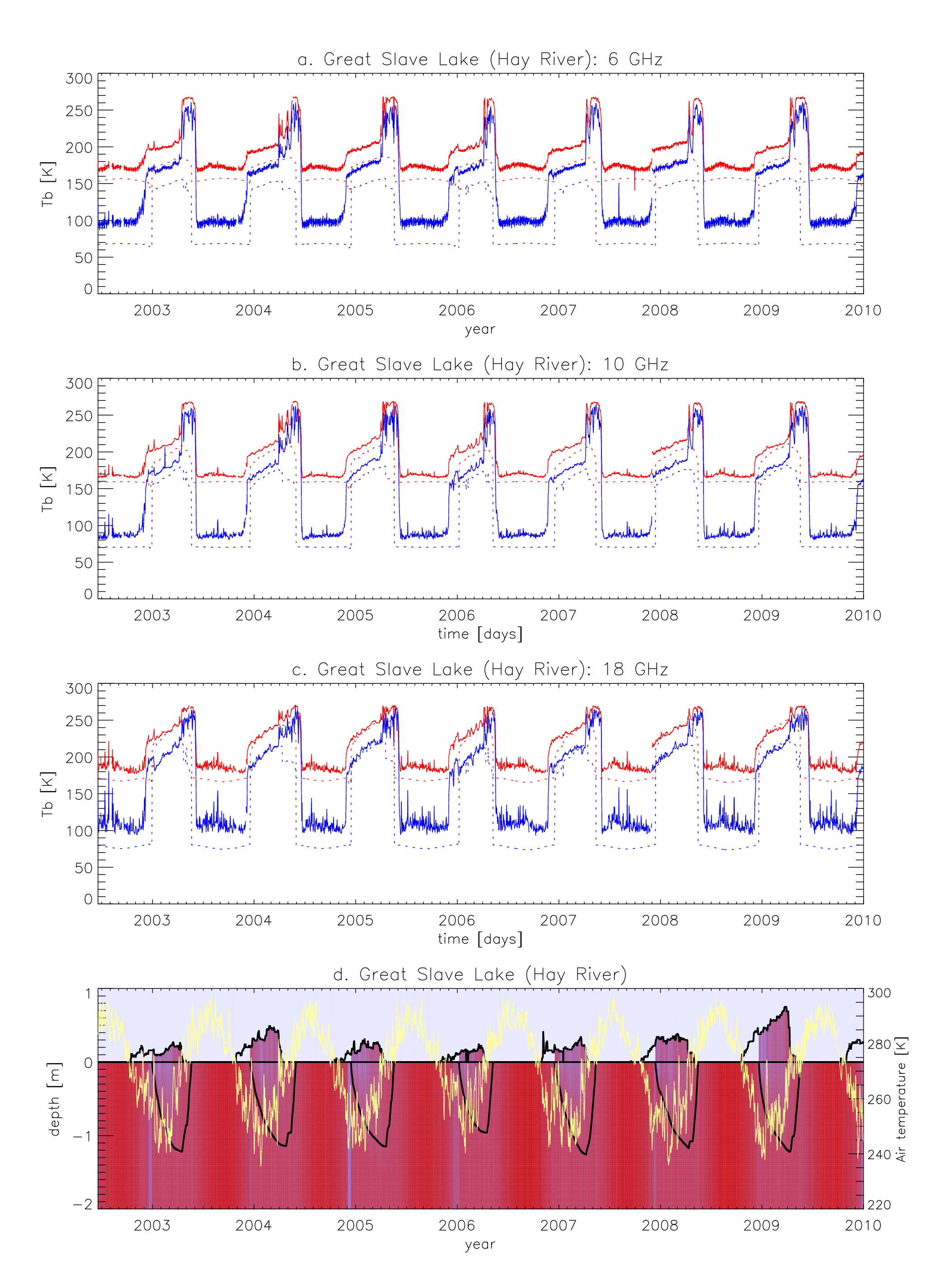}
\includegraphics[width=0.6\textwidth]{GBSL_legend}
\caption{(a), (b) and (c): Time series of measured versus modelled brightness temperatures
over the Great Slave Lake near Hay River, 6, 10 and 18 GHz, respectively.
(d): Ice thickness (black lines), snow depth (black lines), ice and water temperature (shading) and air temperature (yellow line).}
\label{GSL2_tbs}
\end{center}
\end{figure}

The main results are summarized in Figures \ref{GBL_tbs}, \ref{GSL1_tbs} 
and \ref{GSL2_tbs}.
The model shows considerable skill at predicting Tbs over the Great Bear
and Great Slave Lakes, except during the melt season.
Using only temperature, the model has limited predictive power over
open water at 6 GHz.
At 10 GHz, the model shows very little variation over open water, while
at 18 GHz, modelled Tbs are negatively correlated with measured,
showing a lowering of Tb during warm weather where the measured values
are raised.

\begin{figure}
\includegraphics[width=0.9\textwidth]{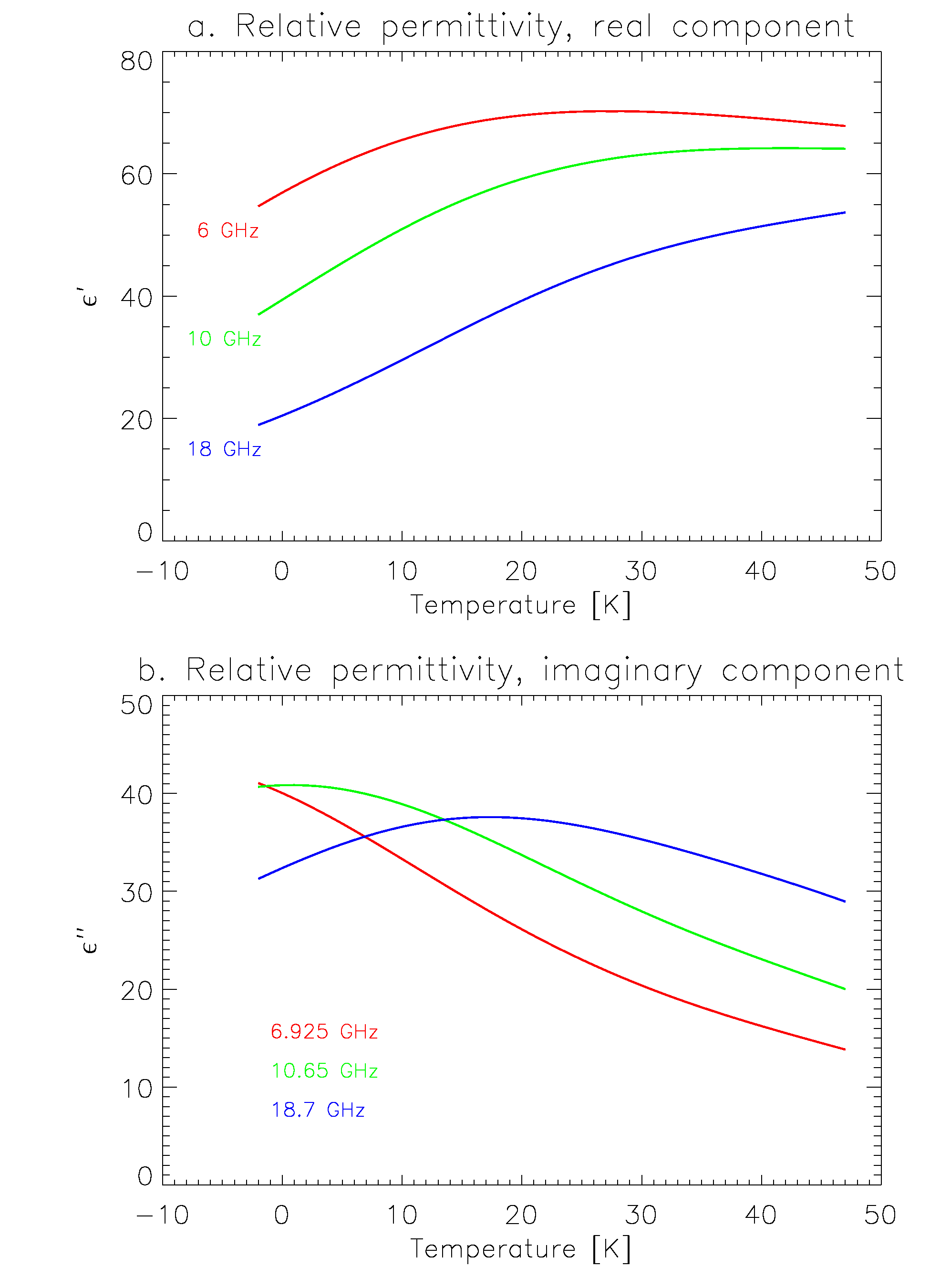}
\caption{Modelled complex relative permittivity of pure water at different
temperatures.}\label{eps}
\end{figure}

The modelled decrease in Tb with increasing temperature over open water at 18 GHz
can be explained by the modelled relative permittivities, shown
in Figure \ref{eps}.
These have been simulated using a Debye relaxation curve \citep{Ulaby_etal1986}
and rise much more steeply at 18 GHz than at the other channels,
although why the resulting Tbs disagree with the actual values is unclear.

\begin{figure}
\includegraphics[width=0.9\textwidth]{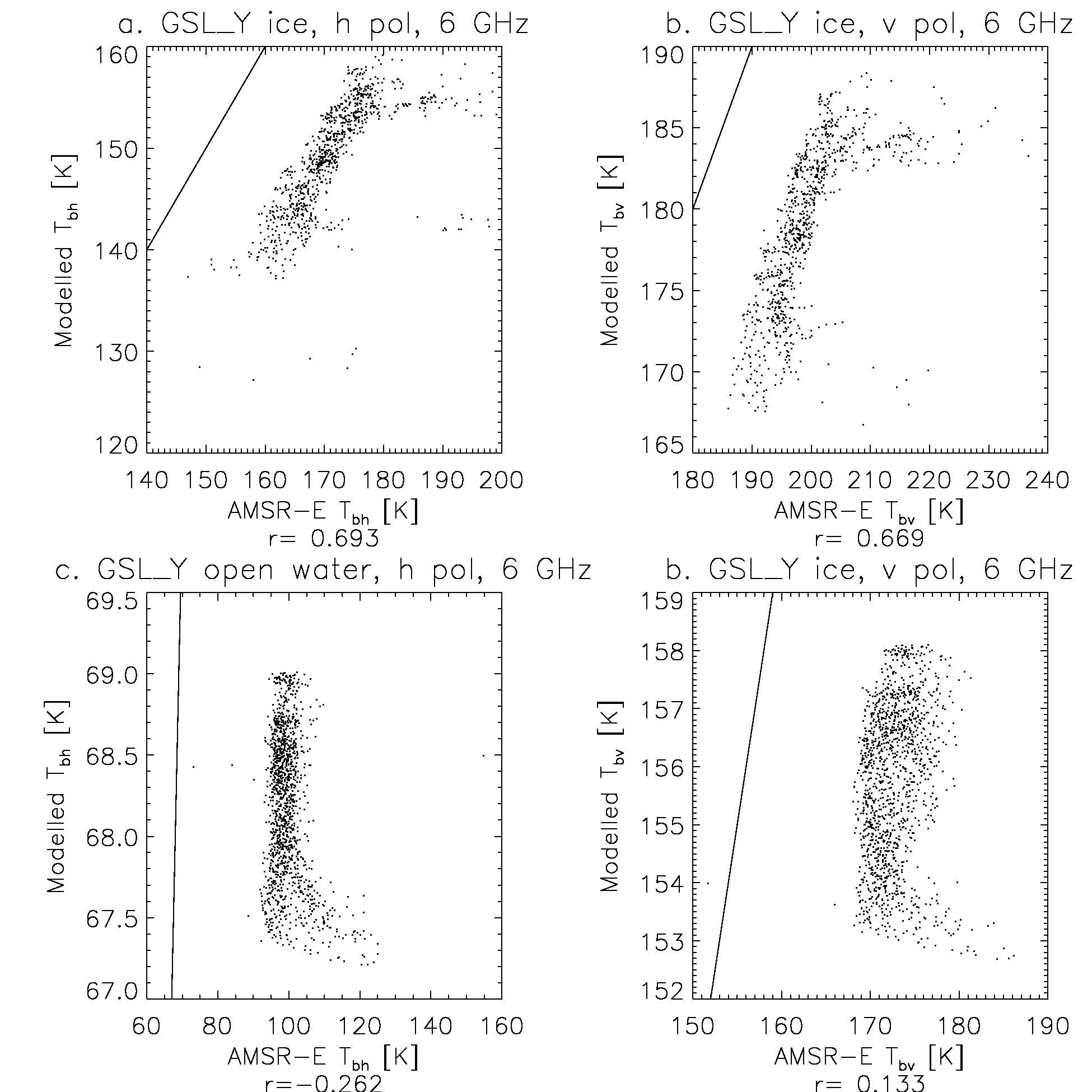}
\caption{Scatterplots of ice during the winter season ((a) and (b), h and v polarizations resp.) and 
open water during the summer months ((c) and (d), h and v polarizations resp.) for 6 GHz.}\label{GSL_Y_scat6}
\end{figure}

\begin{figure}
\includegraphics[width=0.9\textwidth]{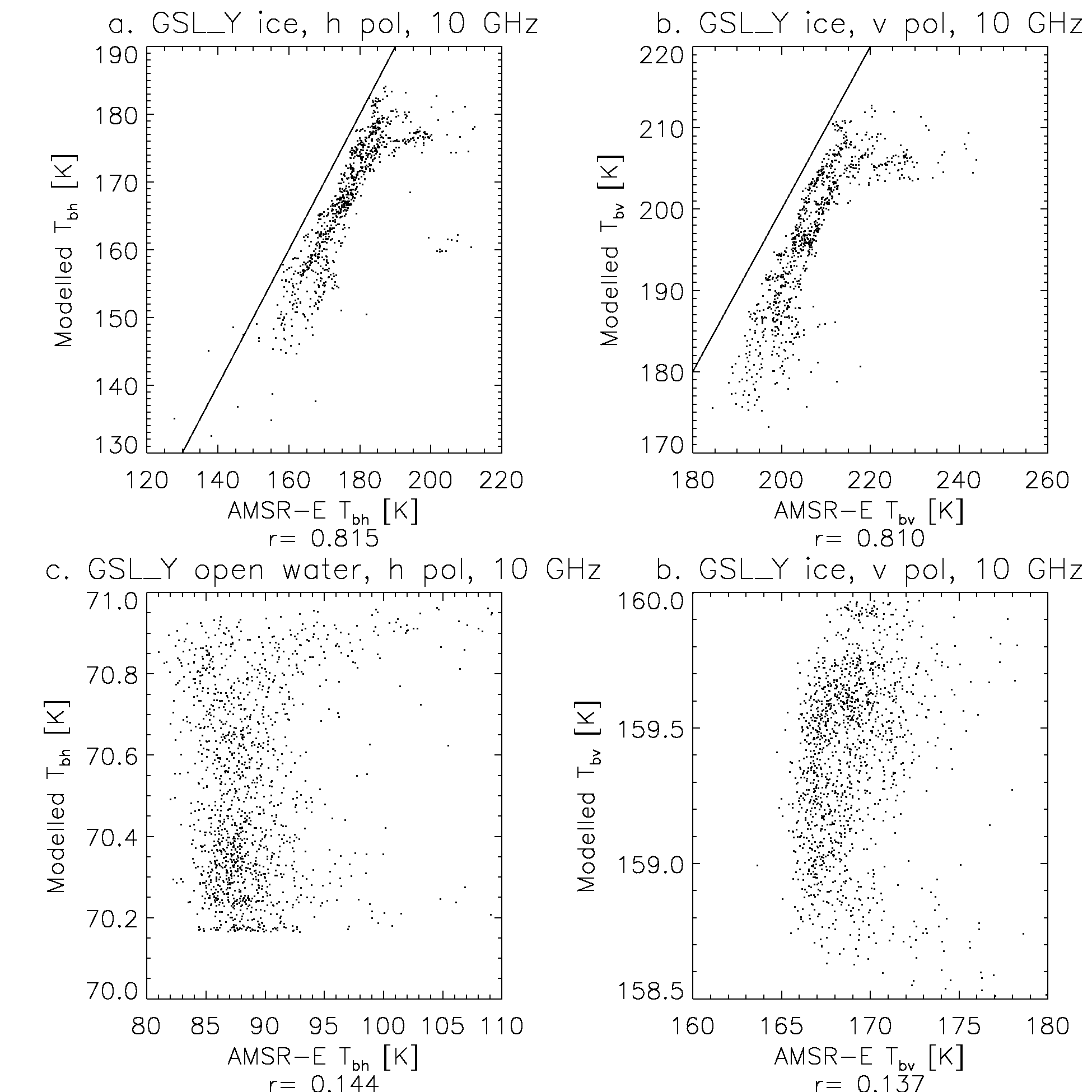}
\caption{Scatterplots of ice during the winter season ((a) and (b), h and v polarizations resp.) and 
open water during the summer months ((c) and (d), h and v polarizations resp.) for 10 GHz.}\label{GSL_Y_scat10}
\end{figure}

\begin{figure}
\includegraphics[width=0.9\textwidth]{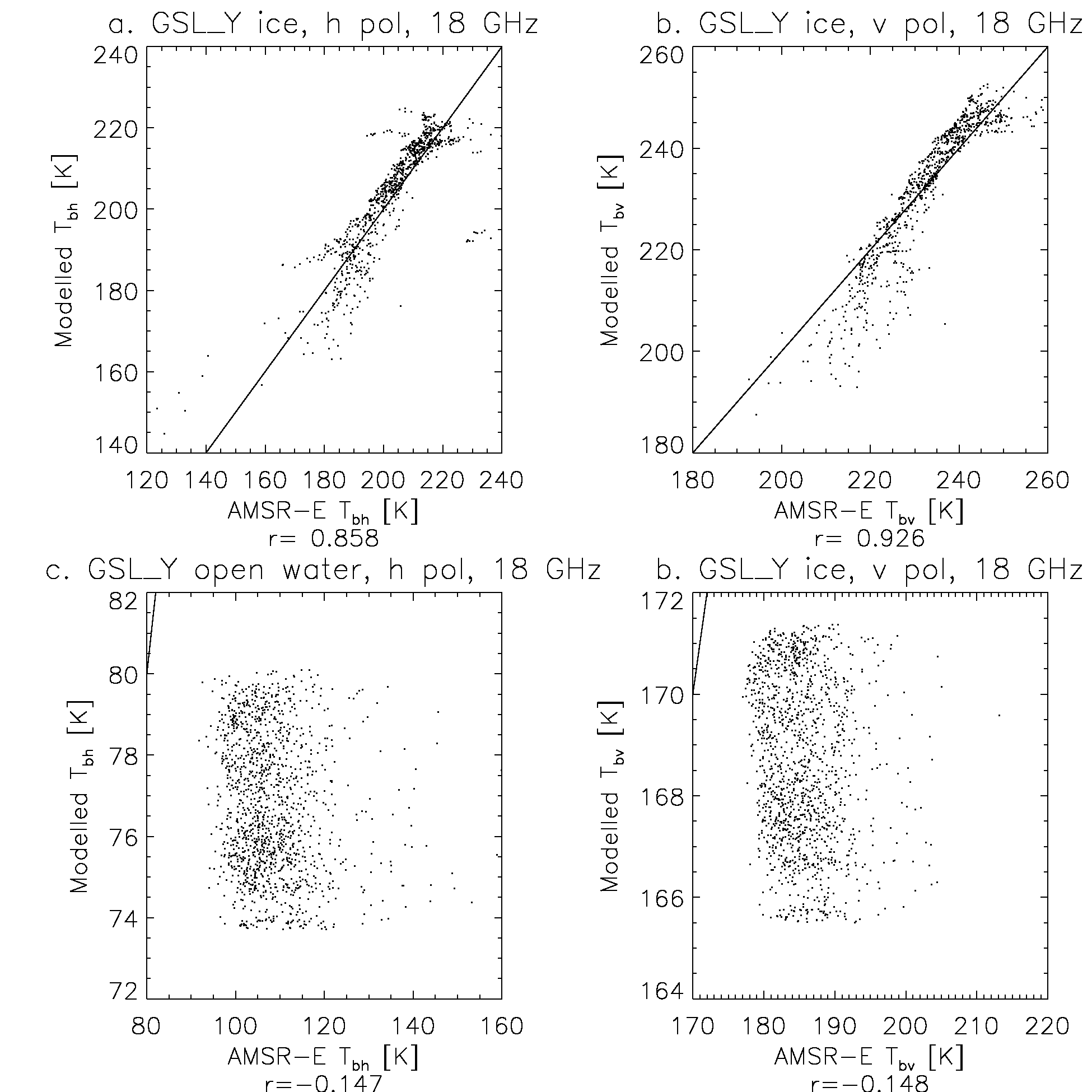}
\caption{Scatterplots of ice during the winter season ((a) and (b), h and v polarizations resp.) and 
open water during the summer months ((c) and (d), h and v polarizations resp.) for 18 GHz.}
\label{GSL_Y_scat18}
\end{figure}

The skill of model predictions over the Great Slave Lake near Yellowknife
for both ice-covered and open water is outlined by the scatterplots in Figures \ref{GSL_Y_scat6},
\ref{GSL_Y_scat10} and \ref{GSL_Y_scat18} for 6, 10 and 18 GHz respectively.
There are two seasons during which the model suffers and both of these have been
excluded from the comparison as they are in some sense beyond the scope
of the model.
Measured Tbs show a marked increase during the melt season, something which
is not reflected in the models.
This is thought to be caused by dirt within the snow which is exposed as it
melts, producing a Tb signature closer to that of soil than of ice or snow.
In \citet{Kang_etal2010}, melt onset is identified by a succession of four
or more days during which the air temperature rises above freezing.
Here we use a more basic technique to select the values of interest,
by simply choosing points where the measured Tbs are between 140 and
200 K at 6 GHz, horizontal polarization.

A similar technique is used to select open water points in which Tbs
are restricted to between 70 and 110 K at 10 GHz, horizontal polarization.
This effectively removes the beginning of fall freeze-up, during which the
lake may be only partially ice-covered 
or contain substantial amounts of
crystallized ice particles (frazil), neither circumstance of which is
encompassed by the ice growth model. 
For a simple method of modelling mixed surface types, see 
\citep{Mills_Heygster2012}. 

Were the model results to be linearly re-calibrated (see below), the residuals
for the vertical polarization over open water, 6 GHz, and over ice, 18 GHz,
would be 2.4 K and 0.78 K, respectively.

\begin{table}
\caption{Brightness temperature biases in Kelvin from the previous study over
Lake Superior for both uncorrected model results and those that have
been corrected for atmospheric effects.  For the ice case, the modelled
values are for 1 m ice thickness.}
\begin{tabular}{|r|r|r|r|r|}
\hline
\multirow{2}{*}{Channel} & \multicolumn{2}{|c|}{Lake Ice} & \multicolumn{2}{|c|}{Open water} \\ 
& uncor. & cor. & uncor. & cor. \\ \hline\hline
  6.925 h & -11.8 & -9.3 & -16.7 & -10.0 \\
\rowcolor[gray]{0.9} 6.925 v & -6.6 &  -5.0 & -7.0 & -3.7 \\
 10.65 h &    -4.5 &   -1.6 &   -18.9 &   -9.5 \\
\rowcolor[gray]{0.9} 10.65 v &     4.8 &    6.5 &   -6.3 &   -1.6 \\
      18.7 h &     8.6 &    15.8 &   -28.0 &    4.3 \\
 \rowcolor[gray]{0.9} 18.7 v &     22.8 &    25.3 &   -9.3 &    6.8 \\
\hline
\end{tabular}\label{superior_bias}
\end{table}

\begin{table}
\caption{Brightness temperature biases in Kelvin for the current study
over the Great Slave Lake near Yellowknife.}
\begin{tabular}{|r|r|r|}
\hline
Channel & Lake ice & Open water \\
\hline
\hline
       6.925 h &  -22.4 &    -31.7 \\
\rowcolor[gray]{0.9} 6.925 v  &    -20.7 &    -17.0 \\
 10.65 h  &    -9.7   &  -18.3 \\
\rowcolor[gray]{0.9} 10.65 v  &    -10.6   &  -9.5 \\
       18.7 h  &     1.0  &   -31.5 \\
\rowcolor[gray]{0.9} 18.7 v &    -0.5   &  -16.9 \\
\hline
\end{tabular}\label{GSL_Y_bias}
\end{table}

While the emissivity model appears to have considerable predictive power,
it is not well calibrated.
Only at 18 GHz do the ice brightness temperatures show relatively little bias.
This is a consistent problem across many of the studies modelling microwave
emissivity conducted by this author 
\citep{SMOSIce_report, Mills_Heygster2009, Mills2012}.
Other studies have also shown biases and other discrepancies with modelled
snow and ice brightness temperatures, depending on the type of model used
\citep{Winebrenner_etal1992,Tedesco_etal2006}.
Possible explanations include calibration problems in the radiometer,
as in \citet{Mills_Heygster2009}, the influence of weather and factors not included
in the model such as scattering, ice ridging, impurities in the ice
and snow cover.
In \citet{Mills2012}, weather was crudely accounted for using a parameterized
correction scheme \citep{Wentz_Meissner2000}.
This step improved the prediction of open water emissivity as seen in
Table \ref{superior_bias} but did little to improve the prediction over
ice as the relative biases
between different frequencies remained similar and in fact became
slightly worse.
Biases over ice were calculated for a modelled ice thickness of 1 m
and uniform, freezing temperature within the ice sheet.

Consider, for comparison, the equivalent biases for the current study,
shown in Table \ref{GSL_Y_bias}.
In both this and the previous study, the spread between the 6 and 8
GHz horizontally-polarized channels is between 23 and 25 K.

\section{Conclusion}

Lake ice brightness temperatures 
over the Great Bear and Great Slave Lakes were simulated 
at 6.925, 10.65 and 18.7 GHz 
using radiative transfer
from ice thicknesses and temperatures 
from the Canadian Lake Ice Model (ClIMo).
These were compared with measurements 
from the Advanced Microwave Scanning Radiometer (AMSR)
on the Earth Observational Satellite (EOS).
Modelled ice Tbs were found to have strong predictive power
when compared with measurements except during the
melt season.
Modelled open water Tbs had some predictive power, but not
during fall freeze-up and not at 10 GHz, while those at 18 GHz
were negatively correlated with measured values.
All Tbs except at 18 GHz over ice, were found to have significant
bias.
The source of this bias is as yet unclear and further work needs
to be done.
The interaction of electro-magnetic radiation with matter is
a fundamental problem in physics and to date, most of the
emissivity models for ice and snowpack are rather ad-hoc,
the current one being no exception.
The study of \citet{Mills_Heygster2009} that investigated the
effect of ice ridging on the microwave signature suggest that
a more rigorous model based on wave interaction rather than
simple line-of-sight would have better predictive power.
It is hoped that future studies can delve more deeply into the
theory and develop more reliable models for ice electromagnetic
properties as well as combine these with more sophisticated
ice growth models.

\section*{Acknowledgements}

Thanks to Kevin Kang and Claude Dugauy for CLIMo results.

\bibliography{fresh_water}

\begin{thebibliography}{}

\bibitem[\protect\astroncite{Cox and Weeks}{1988}]{Cox_Weeks1988}
Cox, G. and Weeks, W. (1988).
\newblock Numerical simulations of the profile properties of undeformed
  first-year sea ice during the gowth season.
\newblock {\em Journal of Geophysical Research}, 93(C10):12499--12460.

\bibitem[\protect\astroncite{Duguay et~al.}{2003}]{Duguay_etal2003}
Duguay, C.~R., Flato, G.~M., Jeffries, M.~O., Menard, P., Morris, K., and
  Rouse, W.~R. (2003).
\newblock Ice-cover variability on shallow lakes at high latitudes: model
  simulations and observations.
\newblock {\em Hydrological Processes}, 17:3465--3483.

\bibitem[\protect\astroncite{Heygster et~al.}{2009}]{SMOSIce_report}
Heygster, G., Hendricks, S., Kaleschke, L., Maass, N., Mills, P., Stammer, D.,
  Tonboe, R.~T., and Haas, C. (2009).
\newblock L-{B}and {R}adiometry for {S}ea-{I}ce {A}pplications.
\newblock Technical Report final report for ESA/ESTEC Contract N.
  21130/08/NL/EL, Institute of Environmental Physics, University of Bremen.

\bibitem[\protect\astroncite{Kang et~al.}{2010}]{Kang_etal2010}
Kang, K.-K., Duguay, C.~R., Howell, S. E.~L., Derksen, C.~P., and Kelly, R.
  E.~J. (2010).
\newblock Sensitivity of {AMSR}-{E} {B}rightness {T}emperature to the
  {S}easonal {E}volution of {L}ake {I}ce {T}hickness.
\newblock {\em IEEE Geoscience and Remote Sensing Letters}, 7(4):751--755.

\bibitem[\protect\astroncite{Mills}{2012}]{Mills2012}
Mills, P. (2012).
\newblock Microwave emissivity of freshwater ice--lake ice and antarctic
  icepack--radiative transfer simulations versus satellite radiances.
\newblock Technical Report arxiv:1202.4095v1, Peteysoft Foundation.

\bibitem[\protect\astroncite{Mills and Heygster}{2011a}]{Mills_Heygster2011}
Mills, P. and Heygster, G. (2011a).
\newblock Retrieving sea ice concentration from smos.
\newblock {\em IEEE Geoscience and Remote Sensing Letters}, 8(2):283--287.

\bibitem[\protect\astroncite{Mills and Heygster}{2011b}]{Mills_Heygster2012}
Mills, P. and Heygster, G. (2011b).
\newblock Sea ice brightness temperature as a function of ice thickness:
  Computed curves for {AMSR}-{E} and {SMOS} (frequencies from 1.4 to 89 {GH}z).
\newblock Technical Report DFG project HE-1746-15, University of Bremen.

\bibitem[\protect\astroncite{Mills and Heygster}{2011c}]{Mills_Heygster2009}
Mills, P. and Heygster, G. (2011c).
\newblock Sea ice emissivity modelling at {L}-band and application to
  {P}ol-{I}ce campaign field data.
\newblock {\em IEEE Transactions on Geoscience and Remote Sensing},
  49(2):612--627.

\bibitem[\protect\astroncite{Sihvola and au~Kong}{1988}]{Sihvola_Kong1988}
Sihvola, A.~H. and au~Kong, J. (1988).
\newblock Effective {P}ermittivity of {D}ielectric {M}ixtures.
\newblock {\em IEEE Transactions on Geoscience and Remote Sensing}, 26(4).

\bibitem[\protect\astroncite{Tedesco et~al.}{2006}]{Tedesco_etal2006}
Tedesco, M., Kim, E.~J., England, A.~W., Roo, R.~D., and Hardy, J.~P. (2006).
\newblock Intercomparison of {E}lectromagnetic {M}odels for {P}assive
  {M}icrowave {R}emote {S}ensing of {S}now.
\newblock {\em IEEE Transactions on Geoscience and Remote Sensing}, 44(10).

\bibitem[\protect\astroncite{Ulaby et~al.}{1986}]{Ulaby_etal1986}
Ulaby, F.~T., Moore, R.~K., and Fung, A.~K., editors (1986).
\newblock {\em Microwave Remote Sensing: Active and Passive, Volume III, From
  Theory to Applications}.
\newblock Artech House, Norwood, MA.

\bibitem[\protect\astroncite{Wentz and Meissner}{2000}]{Wentz_Meissner2000}
Wentz, F.~J. and Meissner, T. (2000).
\newblock {AMSR} {O}cean {A}lgorithm.
\newblock Algorithm Theoretical Basis Document RSS Tech. Proposal 121599A-1,
  Remote Sensing Systems, Santa Rosa CA.
\newblock prepared for NASA Goddard.

\bibitem[\protect\astroncite{Winebrenner et~al.}{1992}]{Winebrenner_etal1992}
Winebrenner, D.~P., Bredow, J., Fung, A.~K., Drinkwater, M.~R., Nghiem, S.,
  Gow, A.~J., Perovich, D.~K., , Grenfell, T.~C., Han, H.~C., Kong, J.~A., Lee,
  J.~K., Mudaliar, S., Onstott, R.~G., Tsang, L., and West, R.~D. (1992).
\newblock Microwave sea ice signature modelling.
\newblock In {\em Microwave Remote Sensing of Sea Ice}, number~68 in
  Geophysical Monographs, chapter~8, pages 137--175. American Geophysical
  Union.

\end{thebibliography}

\end{document}